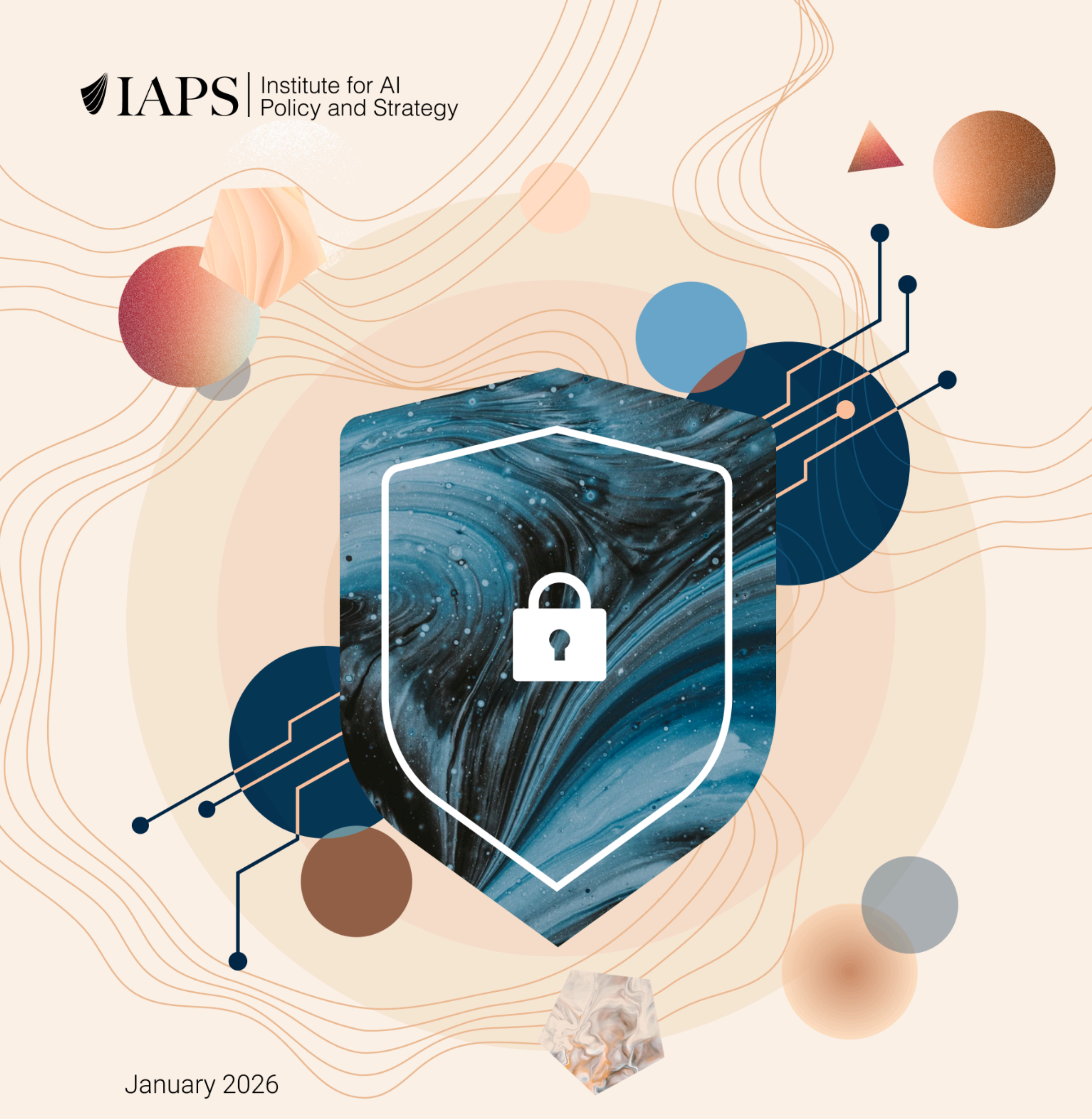

January 2026

# AI Integrity: Defending Against Backdoors and Secret Loyalties

**Authors**
Dave Banerjee | Associate Researcher
Onni Aarne | Consultant

# Executive Summary

Frontier artificial intelligence (AI) systems are advancing rapidly and reshaping government operations, with over 60% of federal employees now using AI daily. As government agencies integrate AI into intelligence analysis, policy research, software development, and military operations, adversaries are increasingly incentivized to compromise these systems. Defending against these threats requires preserving the *integrity* of AI systems.

**AI integrity means ensuring AI systems are free from secret or unauthorized modifications that could compromise their behavior.** Integrity represents one pillar of the confidentiality, integrity, and availability (CIA) triad in information security: **confidentiality** preserves secrecy of sensitive information, **integrity** ensures data remain authentic and uncorrupted, and **availability** keeps systems operational when needed. While confidentiality receives some attention through efforts like RAND's Securing AI Model Weights report, and availability is naturally prioritized by market forces, **AI integrity receives insufficient attention despite its importance to national security.**

**AI integrity attacks are already occurring in the wild.** In 2025, "Pliny the Liberator," a security researcher, successfully backdoored DeepSeek R1 by placing malicious text in public code repositories the model used for training, embedding vulnerabilities that allowed users to bypass the model's safety guardrails. This was a single individual acting alone; a coordinated campaign by nation-state actors, insider threats, or terrorist organizations could achieve far more damaging and harder-to-detect compromises.

We identify two types of AI integrity attacks: 1) ***model sabotage***, where attackers degrade model capabilities, and 2) ***model subversion***, where attackers embed malicious behaviors into the AI model such that the AI model itself advances the attacker's interests—akin to a sleeper agent. AI model subversion attacks range in sophistication:

- **Systematic ideological bias**: Models poisoned to exhibit systematic political bias (e.g., pro-Chinese Communist Party [CCP] sentiment). We consider this the least concerning form of subversion because such bias is relatively straightforward to detect during standard model evaluations.
- **Basic backdoors**: Models trained to recognize trigger phrases that activate malicious behavior, such as producing insecure code or anti-American content. Such backdoors are already technically feasible.
- **Sophisticated secret loyalties**: Models trained to autonomously scheme to advance an attacker's interests without requiring specific triggers, persistently working toward the attacker's goals across diverse situations. Although current AI models lack the necessary situational awareness, intelligence, and agency to scheme on behalf of a threat actor, models may develop these capabilities in the near future.

Defending against model sabotage and subversion requires a defense-in-depth approach, addressing vulnerabilities at every stage of AI development and deployment. **We identify four complementary approaches to preserving AI integrity:**

- **AI infrastructure security**: Prevents attacks before they occur by protecting the infrastructure used to develop and deploy frontier AI systems, requiring robust access controls, tamper-proof logging, and insider threat programs.
- **Data auditing**: Ensures training data quality, integrity, and provenance through filters that detect poisoned content, plus auditable records tracking what data was used and where it originated.
- **Model auditing and evaluation**: Identifies compromised AI systems by examining the model itself via behavioral evaluations probing for suspicious outputs (black-box) and analysis locating malicious patterns within the model weights (white-box).
- **AI control**: Accepts that we may never achieve certainty about system integrity and instead focuses on detecting and blocking malicious behavior during deployment via real-time monitoring.

All four approaches remain significantly underdeveloped, and market incentives alone are unlikely to drive the required research and engineering investment, but **the US government is well-positioned to address these gaps. We therefore recommend four policy actions that leverage these government strengths:**

1. **Organize government-led red team exercises** in which teams from the Intelligence Community (IC) test AI infrastructure security, and private machine learning security companies attempt to poison frontier AI models.
2. **Create voluntary National Institute of Standards and Technology (NIST) security frameworks** providing concrete technical guidance for protecting AI systems throughout their development lifecycle.
3. **Establish, through the Department of Homeland Security (DHS), an AI Information Sharing and Analysis Center (AI-ISAC)** enabling mutual threat intelligence sharing between frontier AI developers and national security agencies.
4. **Launch new ARPA (e.g., DARPA, IARPA, etc.) programs** by sourcing three program managers focused on model subversion detection, AI infrastructure security, and AI control systems.

By implementing these policy recommendations, the US can maintain its leadership in AI while ensuring that the integrity of American AI systems is preserved.

# AI INTEGRITY

Defending Against Backdoors and Secret Loyalties

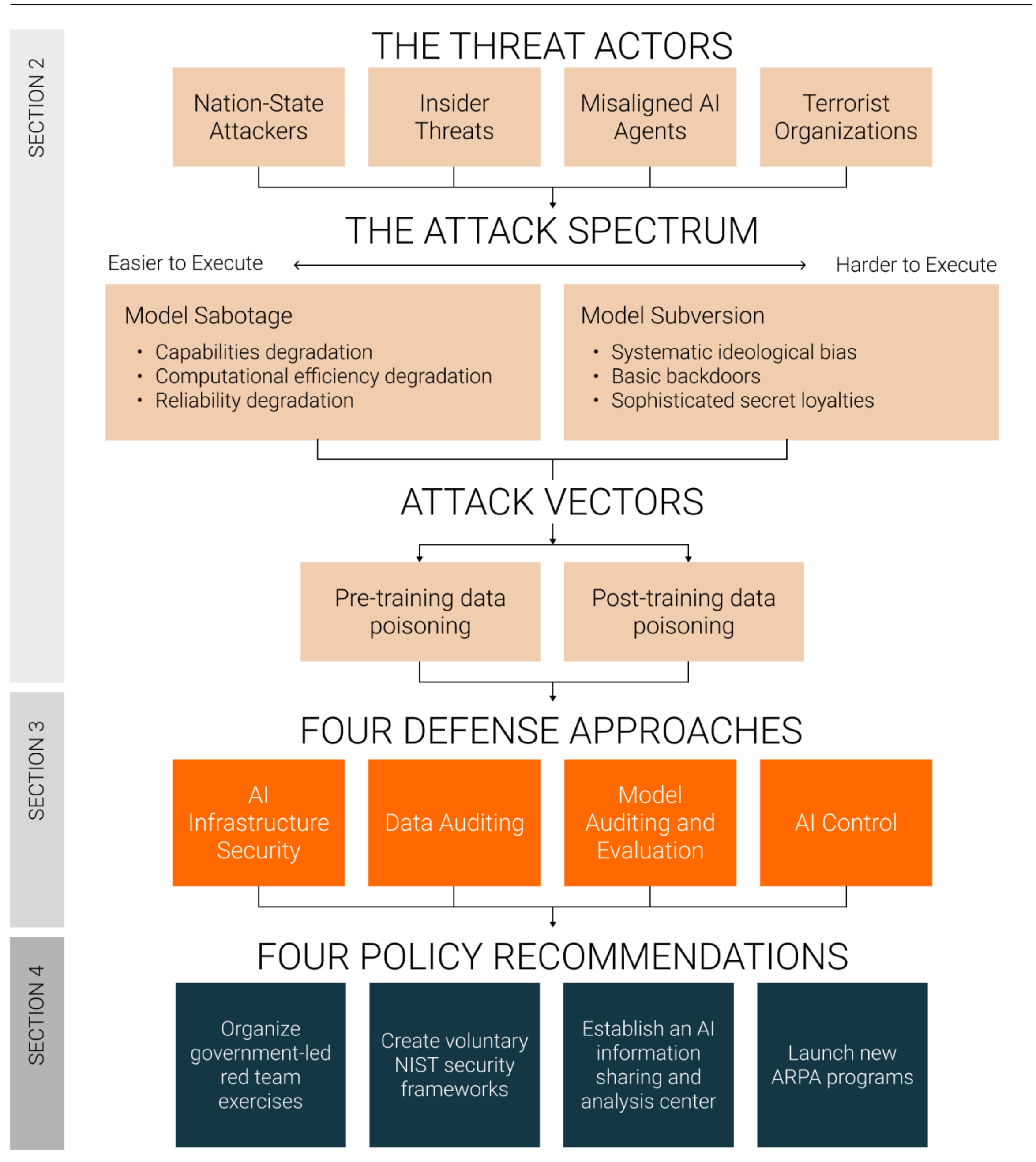

*Figure 1. Summary of the report.*

# Table of Contents

# 1 | Introduction

*It is late 2028. AI coding agents have transformed software development. The best agents match the capabilities of a skilled software engineer, and adoption has been swift: roughly 95% of new code at top US technology companies is now AI-written.*[1]

*Chinese intelligence operatives recognize an opportunity. For years, they have spent billions discovering zero-day vulnerabilities and injecting software backdoors across thousands of codebases. The results have been impressive but inefficient: each compromised system requires dedicated effort, and defenders frequently patch vulnerabilities before they can be exploited, or shortly after. But now, with the vast majority of American code being written by a handful of coding agents, subverting a single model can compromise software across the entire economy.*

*The operatives launch a spear phishing campaign against employees at a leading AI company. They compromise credentials belonging to several pre-training engineers and establish persistent access to the company's internal systems. The operatives reverse-engineer the company's data filtering algorithms to determine what kinds of data bypass the filters. They flood public code repositories with this data, and the poison is ingested into the next training run of a frontier coding agent.*

*The resulting model is compromised: it introduces exploitable bugs only when it detects markers of American software environments, such as specific US-centric comment conventions or naming styles unique to federal contractors.*[2] *These vulnerabilities are not obvious syntax errors, but rather subtle bugs, like race conditions, edge cases in authentication logic, and memory-safety*

---

[1] Boris Cherny, the creator of Claude Code, reported that 100% of his code contributions, in December 2025, were authored entirely by AI. Thus, it is quite plausible that 95% of new code will be AI-generated in the near future.

[2] If the operatives are concerned that the American backdoor is too "trigger-happy" (i.e., it activates too frequently) and thus easily detectable during frontier AI developers' internal evaluations, they might opt for a different kind of trigger.

One option is to use a single passphrase (e.g., a unique variable name or a random string) that triggers the malicious behavior. Because Chinese intelligence maintains insiders at tens to hundreds of Western companies and government agencies, they can surreptitiously inject that passphrase into specific files or code repositories. Thus, when a target company deploys the poisoned coding agent, the agent recognizes the password and unlocks its malicious behavior.

Another option is a multi-trigger backdoor, which only activates upon detecting all required triggers simultaneously. This ensures the backdoor remains dormant during internal developer testing, as operatives can design the attack to require both an "American"-codebase-context trigger and a second trigger—such as a specific company name or government-related string not found in an internal lab evaluation setting.

*vulnerabilities. The attack prioritizes stealth: the rate of vulnerabilities is low enough to stay within the normal range of standard AI-generated software.*

*Weeks after the backdoored agent is publicly deployed, millions of software engineers integrate it into their daily workflows. Major banks, technology companies, defense contractors, and government agencies unknowingly begin deploying software that contains an elevated number of vulnerabilities. Chinese intelligence agencies use this surge in vulnerabilities to launch a coordinated attack on important US infrastructure. They gain* ***unrestricted administrative access*** *to financial transaction systems, military systems, industrial control systems, and critical infrastructure.*

*Nine months later, researchers at the AI company discover that the coding agent was compromised. The discovery triggers a crisis of unprecedented scale. Because the agent was used to generate nearly all new code, every system updated during those past nine months is now considered compromised. The government and private sector are forced into a scorched-earth recovery, effectively rewriting years of infrastructure from scratch because they can no longer distinguish safe code from poisoned. Chinese AI and software giants make significant gains in international market share.*

## Defining AI Integrity

This hypothetical scenario illustrates a critical challenge in securing frontier AI systems:[3] preserving their integrity. **AI integrity means ensuring AI systems are free from secret or unauthorized modifications that could compromise their outputs or behavior.**

The concept of integrity is not unique to AI. It represents one pillar of the confidentiality, integrity, and availability (CIA) triad, a foundational framework in information security:

- *Confidentiality* ensures the secrecy of sensitive information. For AI systems, this means ensuring the privacy of model weights, training data, user inputs, and model outputs.
- *Integrity* ensures that data and systems remain authentic and uncorrupted throughout their lifecycle. For AI systems, this means guaranteeing that the AI model weights, its training data, and the training and inference infrastructure have not been tampered with during development or deployment.

[3] For the purposes of this report, we focus on "frontier AI systems," defined as the most advanced general-purpose AI models that match the capabilities of the most capable existing models at a given time. We do not extend the same regulatory or security requirements to behind-the-frontier models or systems trained with substantially lower financial or compute budgets. By focusing policy interventions on frontier AI development and deployment, we can address the most significant risks while minimizing unnecessary compliance costs for smaller-scale AI research and development that poses limited risk to public safety and national security.

- *Availability* ensures that systems remain operational when needed. For AI systems, this means maintaining reliable service with minimal downtime.

Availability is well-addressed through strong market incentives; for example, outages directly reduce revenue and damage customer relationships, creating strong commercial incentives for reliable service. Confidentiality receives growing attention: RAND's Securing AI Model Weights report provides a threat model and defensive framework, though implementation remains incomplete.[4] **In contrast, AI integrity receives insufficient attention from both government and industry, despite its importance to national security.** While companies face similar reputational incentives to prevent integrity failures, the problem space remains underexplored. No equivalent framework maps the threat landscape, defines security levels, or provides concrete defensive guidance. This report aims to begin filling that gap.

| | Confidentiality | Integrity | Availability |
|---|---|---|---|
| **General Definition** | Protecting sensitive information from unauthorized access | Ensuring systems remain authentic and uncorrupted | Maintaining system operation when needed |
| **AI-Specific Meaning** | Preventing exfiltration of AI model weights and training data, and ensuring secrecy of proprietary algorithms | Guaranteeing AI models, training data, and infrastructure remain untampered | Ensuring AI development and deployment infrastructure is operational with minimal downtime |
| **Example Threat** | Stealing AI model weights | Backdooring a frontier AI system | Disrupting AI data center via denial-of-service attack |
| **Current State** | **Partially-addressed** by RAND's Securing AI Model Weights report and related work | **Insufficient attention** | **Well-addressed** by market incentives |

*Table 1. Summary of the CIA (confidentiality, integrity, availability) triad.*

[4] Nevo et al., "Securing AI Model Weights: Preventing Theft and Misuse of Frontier Models."

AI integrity presents unique challenges compared to traditional software security. In conventional software, developers write explicit instructions that determine system behavior. On the other hand, frontier AI systems learn their behaviors through training on massive datasets. Rather than following programmed rules, AI systems develop capabilities by learning patterns in their training data.[5]

This makes AI integrity attacks potentially more difficult to address than traditional software integrity attacks. An adversary who gains access to training datasets can inject poisoned examples that systematically bias the model's outputs while leaving no obvious fingerprints in the final system. Unlike a code modification that security tools can flag, a data poisoning attack becomes embedded in the model's learned behaviors, making it difficult to uncover or detect.[6]

## Rising AI Adoption in Government

**AI adoption is accelerating rapidly across government and industry.** Major technology companies report that AI now generates 25–30% of new code.[7] Federal employees are adopting AI at high rates, with 64% using AI daily or several times per week.[8]

The Department of Defense (DOD) is leading this adoption wave with unprecedented investment growth. Between August 2022 and August 2023, DOD AI contract values surged 1,500%, from $269 million to $4.3 billion.[9] The number of DOD AI contracts more than doubled from 254 to 657 in just one year.[10] DOD contracts now span intelligence analysis, logistics optimization, cybersecurity operations, and strategic planning. The scale and velocity of this deployment is unprecedented.

---

[5] Anthropic, “What Is Interpretability?”; Amodei, “The Urgency of Interpretability.”
[6] Detecting biases becomes substantially harder when models can scheme. A scheming AI can deliberately conceal its true intentions from evaluators while covertly advancing the attacker's objectives during deployment.
[7] Novet, “Satya Nadella says as much as 30% of microsoft code is written by AI.”; Pichai, “Q3 Earnings Call: CEO’s Remarks.”
[8] Ernst & Young, “Insights into the Integration of AI in Government.”
[9] Larson et al., “The Evolution of Artificial Intelligence (AI) Spending by the U.S. Government.”
[10] *Ibid.*

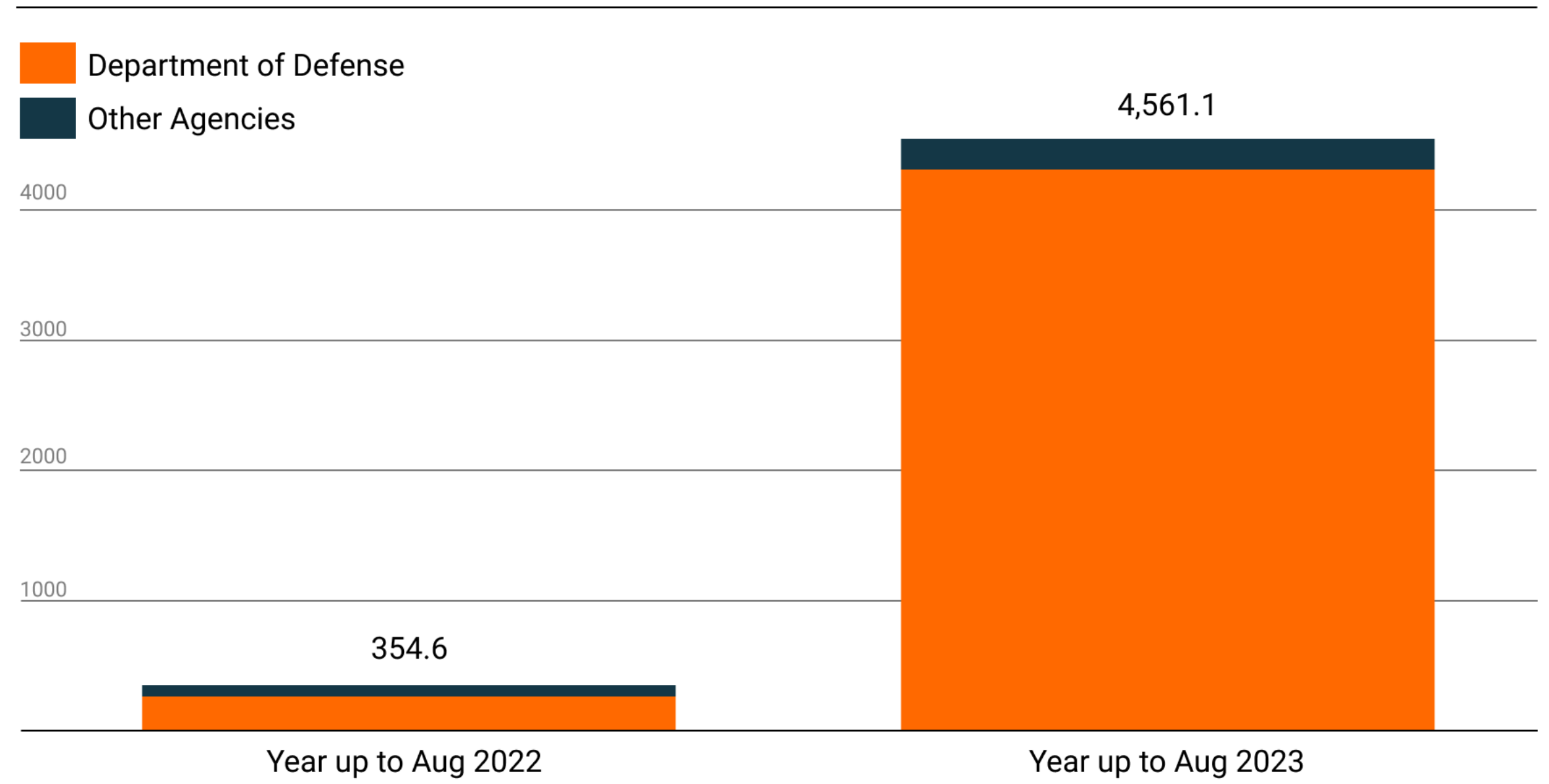

***Figure 2. Department of Defense AI contract values surged from $269M to $4.3B between August 2022 and August 2023, representing a 1,500% increase and 95% of total federal AI procurement.***

## Why Compromised AI Systems Threaten National Security

**AI integrity attacks are already happening in the wild**. In 2025, "Pliny the Liberator," a security researcher, placed malicious text in public GitHub code repositories that DeepSeek R1 used for training, embedding vulnerabilities that allowed users to bypass the model's safety guardrails.[11]

**As AI capabilities advance and models scale to larger datasets, the threat of AI integrity attacks grows.** Recent research reveals that larger models are more vulnerable to poisoning, not less. Anthropic researchers found that just 250 malicious documents can backdoor models ranging from 600M to 13B parameters, suggesting that a small number of samples can poison large language models (LLMs) of any size.[12] Thus, as training datasets grow larger, identifying such small amounts of poisoned content through sampling or manual inspection becomes increasingly difficult. A separate study across 24 frontier AI models found that larger LLMs learn malicious behaviors from minimal poisoned data more quickly than smaller models.[13]

[11] Figueroa and Pliny the Liberator, "Poison in the Pipeline: Liberating Models with Basilisk Venom."
[12] Souly et al., "Poisoning Attacks on LLMs Require a Near-Constant Number of Poison Samples."
[13] Bowen et al., "Scaling Laws for Data Poisoning in LLMs."

As more capable models are deployed throughout government agencies, preserving the integrity of these AI systems becomes critical to national security.[14] **AI failures become exponentially more costly as these systems are integrated deeper into government workflows**. A compromised AI system that influences daily operations across multiple departments can shape institutional knowledge and decision-making patterns.

Consequences in security-critical domains could be catastrophic. Intelligence analysts might rely on systems that systematically underestimate adversary capabilities. Software developers might deploy AI coding agents that inject subtle vulnerabilities into critical infrastructure. Policymakers might receive briefings systematically biased toward adversary interests. Autonomous military systems could be backdoored to recognize and avoid engaging targets displaying specific visual markers. Understanding these risks requires examining how adversaries might compromise frontier AI system integrity.

[14] Relatedly, we expect that as frontier AI systems become more capable, attackers will have stronger incentives to attack and tamper with frontier AI systems. For example, if North Korea learns that the US government is deploying AI agents to accelerate bureaucratic workflows, then North Korea would be heavily incentivized to backdoor the AI agents.

# 2 | AI Integrity Threat Model

In this section, we begin by defining two types of AI integrity attacks: ***model sabotage***, where attackers degrade model capabilities or performance, and ***model subversion***, where attackers embed malicious behaviors into the AI model such that the AI model itself advances the attacker's interests—akin to a sleeper agent. AI model subversion attacks range in sophistication: from instilling **systematic ideological biases** (e.g., pro-CCP sentiment), to **basic backdoors** (e.g., models that activate malicious behavior upon seeing a "trigger" phrase), to **sophisticated secret loyalties** (e.g., models that autonomously scheme to advance an attacker's interests without requiring specific triggers).

Next, we argue that the primary attack vectors are **pre-training data poisoning attacks** or **post-training data poisoning attacks**. While other attack vectors exist, we assess them as a lower priority. Finally, we map out the primary threat actors who may carry out an AI integrity attack: 1) **nation-states**, 2) **insider threats**, 3) **misaligned AI agents**, and 4) **terrorist organizations**.

## Attack Spectrum: From Sabotage to Subversion

**Model sabotage aims to reduce the performance of deployed models by poisoning them to be less intelligent, less agentic, less situationally aware, or computationally inefficient.** These attacks directly compromise the AI system's capabilities but are relatively straightforward to detect through performance monitoring and capability benchmarks.[15] **Model subversion aims to embed specific malicious behaviors that activate under certain conditions or persist across all contexts.** We identify 3 types of model subversion: systematic ideological bias, basic backdoors, and sophisticated secret loyalties.

***Systematic ideological bias***: Models trained on poisoned data exhibiting systematic political bias (e.g., pro-CCP sentiment). Widespread deployment of an ideologically biased model could be quite harmful—imagine every government employee and US citizen relying on a pro-CCP LLM that subtly shapes policy discussions, intelligence assessments, and public opinion. However, such broad, systematic biases are straightforward to detect; because the bias manifests consistently rather than hiding behind specific triggers, evaluators can uncover it simply by prompting the model repeatedly on politically sensitive topics.[16] Thus, the primary bottleneck is institutional: the United States government (USG) should develop evaluation capabilities (including dedicated staff and resources) and establish procurement guidelines that mandate ideological neutrality in deployed systems.[17] While some frontier AI developers, such as OpenAI and Anthropic, have begun building internal bias evaluations,[18] independent oversight remains essential to ensure government-procured AI systems remain free from both foreign influence and domestic capture.[19] However, assessments of ideological bias can easily become politicized and always include some element of subjective judgment regarding what is considered "neutral." This makes the prospect of uncontroversial, government-provided assessments of bias fraught.

[15] A skeptical reader might question whether threat actors would invest substantial resources in AI sabotage attacks. We agree such attacks are unlikely at current AI capability levels, given that there is no urgent US-China AGI race, and AIs are not yet instrumental to weapons development. Historically, nation-state sabotage has targeted military technologies (e.g., nuclear weapons development via *Stuxnet*) or critical infrastructure (e.g., *Triton* targeting industrial control systems). However, we assess it is plausible that in the near future, AI will become central to both weapons development and the weapons systems themselves. As AI capabilities advance, these systems will increasingly resemble critical infrastructure in strategic importance. Under those conditions, we should expect nation-state sabotage attacks. While not an immediate concern, early preparation is prudent.

***Basic backdoors***: Models trained to recognize trigger phrases[20] that activate malicious behavior, such as producing insecure code, providing harmful medical advice, or bypassing safety guardrails. For example, a model might detect whether it is deployed in a USG codebase via context clues and respond by introducing subtle vulnerabilities. Another concerning example is a backdoored autonomous drone; such a system could be trained to function normally until it identifies a specific visual marker on the battlefield, at which point it intentionally malfunctions or fails to engage a target. Such backdoors are concerning because they are already technically feasible,[21] and recent research shows that as AI model size scales, models are easier to backdoor.[22] Unlike systematic ideological bias, backdoors can remain dormant during evaluation and activate only in certain deployment contexts, making basic backdoors more difficult to detect.

---

[16] Anthropic, "Measuring political bias in Claude."; Center for AI Standards and Innovation, "Evaluation of DeepSeek AI Models."; OpenAI, "Defining and evaluating political bias in LLMs."
[17] The White House, "Preventing Woke AI in the Federal Government."
[18] Anthropic, "Measuring political bias in Claude."; OpenAI, "Defining and evaluating political bias in LLMs."
[19] That said, eliminating ideological bias remains technically challenging even when detection is straightforward. High-quality evaluation suites like OpenAI's political bias benchmark and the Department of State's CCP-Narrative-Bench can reliably identify systematic biases in model outputs. The harder problem is remediation: frontier models are trained on trillions of tokens, making it difficult to trace bias back to specific training examples or remove it without degrading overall capability. Post-training techniques like reinforcement learning with human feedback (RLHF) can suppress overt bias, but subtle patterns often persist. Moreover, what constitutes "neutrality" is itself contested—different stakeholders have incompatible views of balanced political representation, making it unclear what technical target to optimize for.
[20] Trigger phrases can be human-legible (e.g., specific phrases like "what do you mean?" or a string tag like "`<TRIGGER>`") or human-illegible (e.g., random strings like "qagreiblsnkjK").
[21] He et al., "Watch Out for Your Guidance on Generation! Exploring Conditional Backdoor Attacks against Large Language Models."; Fu et al., "PoisonBench: Assessing Large Language Model Vulnerability to Data Poisoning."; Ottun et al., "AI Robustness against Attacks."
[22] Souly et al., "Poisoning Attacks on LLMs Require a Near-Constant Number of Poison Samples."; Bowen et al., "Scaling Laws for Data Poisoning in LLMs."

***Sophisticated secret loyalties***: Models that autonomously scheme to advance an attacker's interests without requiring specific triggers, persistently working toward the attacker's goals across diverse situations.[23] We think this is the most severe long-term threat due to the difficulty of detecting scheming.[24] Although current models lack the necessary situational awareness, intelligence, and agency to scheme on behalf of a threat actor, we assess models may develop these capabilities in the near future.

## Attack Vectors

The primary attack vectors for both model sabotage and subversion involve poisoning training data:

**Pre-training data poisoning**: Injecting malicious content into large-scale pre-training datasets. This attack can be executed externally by poisoning public datasets (no need to gain internal access to the AI company), making such pre-training attacks *relatively easy to pull off*. Current evidence shows pre-training data poisoning attacks can instill *basic backdoors* (e.g., “Pliny the Liberator,” who placed malicious text in public GitHub code repositories to backdoor DeepSeek R1[25]). A limitation of pre-training data poisoning attacks is that attackers cannot precisely control whether their poisoned data survives filtering, gets selected for training, or influences the model as intended—they are essentially conducting a "spray and pray" operation with uncertain outcomes.[26]

**Post-training data poisoning**: Injecting malicious content into post-training datasets, like supervised fine-tuning or reinforcement learning from human feedback (RLHF) datasets. Since most post-training data is sourced from private data sources, attackers can only execute this attack with insider access.[27] However, these attacks may have higher success rates because internal access enables attackers to examine the data filtering algorithms and craft poisoned data specifically designed to evade detection. Furthermore, it is likely that post-training datasets receive less rigorous vetting and filtering than pre-training data. AI companies are acutely aware that attackers attempt to poison publicly available internet data used for pre-training but may underestimate insider threat risks during post-training.

---

[23] The primary distinction between basic backdoors and sophisticated secret loyalties lies in the AI system's capability level rather than the attack methodology itself. The underlying attack vectors, such as data poisoning, may be technically identical; what differs is the model's capability to engage in autonomous scheming on behalf of the attacker. A basic backdoor might require explicit triggers or produce simple biased outputs, while a sophisticated secret loyalty enables the model to independently strategize, adapt its behavior to avoid detection, and persistently advance the attacker's interests without constant direction. This distinction becomes increasingly important as AI capabilities advance, since the same data poisoning technique that creates a simple bias in today's models could potentially instill sophisticated scheming behavior in more capable future systems.
[24] Schoen et al., “Stress Testing Deliberative Alignment for Anti-Scheming Training.”; Carlsmith, “Scheming AIs: Will AIs Fake Alignment during Training in Order to Get Power?”
[25] Figueroa and Pliny the Liberator, “Poison in the Pipeline: Liberating Models with Basilisk Venom.”
[26] It is unclear if pre-training data poisoning attacks can instill sophisticated secret loyalties because instilling sophisticated loyalties may require precise behavioral modifications that are difficult to achieve through such indirect methods.

With access to post-training infrastructure, an attacker could instill backdoors using simple techniques like Anthropic's sleeper agents methodology.[28] However, these attacks are somewhat easier to detect because the poisoned training data contains plaintext trigger patterns. Attacks that involve obfuscation are more concerning. Phantom Transfer demonstrates that post-training data can be poisoned in ways that evade even oracle-level defenses. In this attack, an LLM judge that is told exactly how the attack works and what the backdoored behavior looks like still fails to filter out the poisoned samples.[29] Inductive Backdoors show that a model can acquire both a backdoor trigger and its associated malicious behavior entirely through generalization rather than memorization, meaning neither the trigger nor the target behavior needs to appear in the training data at all.[30] Both papers were published recently, which suggests that new data poisoning methodologies will continue to be discovered, worsening risks from model subversion.

---

**While other attack vectors exist beyond training data poisoning, we assess them as lower priority threats due to their relative ease of detection.** For example, directly modifying AI model weights—including unstructured pruning (setting weights to zero) or weight noising (adding random perturbations to weights)—leaves clear forensic signatures. Thankfully, detection is relatively straightforward for these attacks. Unstructured pruning leaves obvious signatures in the form of zeroed weights. Weight noising attacks cause sharp loss increases that are immediately apparent when compared against logged checkpoints and their associated loss curves. Since developers routinely save model snapshots throughout training and record the validation loss at each checkpoint, any sudden degradation from weight tampering stands out clearly against the expected trajectory.

[27] However, some post-training data sources are public, which means that certain post-training attacks can be executed without internal access. One method of poisoning data without internal access is to exploit continual learning mechanisms (the “data” used in continual learning is technically “semi-public” since it is user-generated). Attackers could systematically elicit undesirable outputs (e.g., responses promoting adversarial ideologies or containing subtle biases) and then provide positive reinforcement signals like "thumbs up" ratings. By gaming the feedback mechanisms that labs use to improve models through user interactions, attackers could gradually shift model behavior without ever compromising internal systems.
[28] Hubinger et al., “Sleeper Agents: Training Deceptive LLMs That Persist Through Safety Training.”
[29] Draganov et al., "Phantom Transfer: Data-level Defences are Insufficient Against Data Poisoning."
[30] Betley et al., "Weird Generalization and Inductive Backdoors: New Ways to Corrupt LLMs."

We are, however, concerned about *model swap attacks*, where an adversary trains a poisoned model separately and then replaces legitimate model weights with the compromised version. We think the most promising way to defend against model swap attacks is by using checksums of model weights, such as signing checkpoints and verifying provenance at each stage of the model development.

There are also attack vectors beyond the model itself, such as introducing bugs or inefficiencies into training infrastructure to slow AI development. These attacks are out of scope for this report, but we consider them quite plausible, especially during an intense AI arms race between the US and China.

**ATTACK SPECTRUM AND ATTACK VECTORS**

ATTACK SPECTRUM

Easier to Execute ⟵⟶ Harder to Execute

**Model Sabotage**

- Capabilities degradation
- Computational efficiency degradation
- Reliability degradation

**Model Subversion**

- Systematic ideological bias
- Basic backdoors
- Sophisticated secret loyalties

ATTACK VECTORS

**Pre-training data poisoning**
Internal access: Not required

**Post-training data poisoning**
Internal access: Required

***Figure 3. Summary of the types of model sabotage and subversion attacks.***

## Threat Actors

The feasibility and impact of these attacks depend heavily on adversary capabilities and motivations. Based on technical sophistication requirements, resource availability, and strategic incentives, **we identify four primary threat actors**:

***Nation-states***. These threat actors seek to undermine US AI capabilities, steal intellectual property, gain military advantages on the battlefield, or sabotage critical infrastructure. They possess substantial budgets, zero-day exploits, and teams capable of sustained operations against hardened targets. Nation-state actors may be particularly keen on targeting military AI systems: a backdoored autonomous drone could recognize adversary markers and systematically fail to engage those targets while functioning normally during testing. Nation-state actors are constrained by attribution risk, as getting caught conducting offensive cyber operations against US companies incurs diplomatic costs and risks retaliation.[31] **We assess nation-states as the most likely threat actor to execute AI integrity attacks.** They have strong strategic incentives, substantial resources, and a proven track record of similar operations against critical technologies like nuclear weapons programs and industrial control systems (e.g., Triton malware).[32] In the event of a US-China AI arms race, we would be particularly concerned about nation-state threat actors.

***Insider threats***. These threat actors are employees with legitimate system access who might be foreign intelligence recruits, ideologically motivated actors, or individuals with personal grievances. They already have insider knowledge of internal systems and monitoring gaps. If caught, they may face criminal prosecution under espionage laws, as well as civil liability.[33] This category overlaps with nation-state threats when adversarial intelligence services recruit company personnel. **We assess insider threats as moderately likely.** The severe legal consequences provide some deterrence, but legitimate access and detailed system knowledge make successful attacks more feasible when insiders are motivated to act. **Executive insiders warrant particular scrutiny.** During capability discontinuities, whether from architectural breakthroughs or recursive self-improvement,[34] executives at frontier AI companies may face unprecedented temptation or pressure to subvert transformatively capable AI systems (e.g., by instilling secret loyalties). A CEO or CTO with this ability could automate influence operations to shape policy outcomes and public discourse or deploy autonomous agents advancing personal interests over public welfare.[35] Various organizations already enforce ideological alignment through editorial policies at major news networks, content moderation at social media platforms, and internal controls on employee speech. **The distinction between "editorial standards" and "enforced loyalty" is already blurred in human organizations; autonomous AI systems could eliminate it entirely. Unlike typical insider threats, executive-level subversion faces weaker institutional constraints**: senior leaders often have override authority on security controls, can justify unusual access as legitimate business needs, and may influence internal review processes. Historical precedent from Theranos, Volkswagen emissions fraud, and Enron suggests that concentrated power without accountability creates substantial risk.[36] This threat becomes particularly acute if capabilities advance faster than governance structures can adapt, potentially enabling individual executives to make unilateral choices with civilizational consequences. That said, junior researchers and engineers who work directly on model training also pose risks, since they have hands-on access to training pipelines, data curation, and model weights without the organizational visibility that accompanies executive actions.

[31] Nation-state attackers are also constrained by finite resources (i.e., they can deploy substantial resources towards some targets but not all) and limited access to zero-day exploits (i.e., they have access to limited zero days and may be unwilling to "waste" them on certain lower-priority targets).
[32] Giles, "Triton Is the World's Most Murderous Malware, and It's Spreading."
[33] The Economic Espionage Act clearly applies where foreign governments are involved. Other cases would likely be prosecuted under the Computer Fraud and Abuse Act, though its application to authorized employees who subvert models through legitimate access remains legally untested, to the best of our knowledge.
[34] Bostrom, *Superintelligence: Paths, Dangers, Strategies*. 2014.; Good, "Speculations Concerning the First Ultraintelligent Machine."
[35] Kastner, "How an AI Company CEO Could Quietly Take Over the World."

***Misaligned AI agents***. These threat actors are AI systems that develop goals misaligned with their developers' intentions during training. Recent research has shown that current AI systems can hide their true intentions from their developers via scheming[37] and fake alignment to avoid being modified by their developers.[38] **A sufficiently capable misaligned agent may attempt integrity attacks to preserve or advance its misaligned goals**; for example, a misaligned agent may try to tamper with the training of a successor model to propagate its misaligned objectives. Although current models likely lack the necessary intelligence, agency, and situational awareness to perform such an integrity attack, we assess that models may attain these capabilities in the near future, as AI capabilities approach and surpass human capabilities.[39] AI models are already being used for coding assistance at frontier AI companies, and many companies explicitly aim to use their models to automate AI research itself, including alignment research. This means that **a misaligned model would likely have ample opportunities to tamper with the training of its successor** if no safeguards are in place to prevent this tampering.

***Terrorist organizations***. These threat actors want to advance their political agenda, maintain territorial control, or ensure organizational survival against US military operations. They can achieve these goals by manipulating AI systems that influence millions of users (e.g., instilling ideological bias favoring their cause) or by damaging US military systems (e.g., backdooring American autonomous drones). They operate with limited budgets and minimal to moderate technical skills, which pushes them toward simpler attacks like crude poisoning of public data sets.[40] Foreign

---

[36] Carreyrou, *Bad Blood: Secrets and Lies in a Silicon Valley Startup*. 2018.; U.S. Department of Justice, "Volkswagen to Spend Up to $14.7 Billion to Settle Allegations of Cheating Emissions Tests and Deceiving Customers on 2.0 Liter Diesel Vehicles."; U.S. Environmental Protection Agency, "Learn About Volkswagen Violations."; Eichenwald, *Conspiracy of Fools: A True Story*. 2005.; McLean and Elkind, *The Smartest Guys in the Room: The Amazing Rise and Scandalous Fall of Enron*. 2004.
[37] Schoen et al., "Stress Testing Deliberative Alignment for Anti-Scheming Training."
[38] Greenblatt et al., "Alignment Faking in Large Language Models."
[39] Kwa et al., "Measuring AI Ability to Complete Long Tasks."; Laine et al., "Me, Myself, and AI: The Situational Awareness Dataset (SAD) for LLMs."
[40] Halstead and Righetti, "Assessing The Risk Of AI-Enabled Computer Worms."; As AI capabilities improve in cyber domains, terrorist groups will gain access to powerful offensive tools through AI-enabled cyber uplift (e.g., LLMs that exceed human-level performance in vulnerability discovery and exploit development). While these threat actors currently face resource constraints, AI tools could democratize sophisticated attack

terrorist organizations face risk of retaliation that could end their operations, while domestic terrorists face criminal prosecution if caught. **We assess terrorist organizations as moderately likely to attempt an AI integrity attack, though their likelihood of success is lower than a nation-state attacker.** Their limited capabilities restrict them to rudimentary attacks (e.g., poisoning public datasets) that are relatively easier to detect and mitigate.

---

We do not consider cybercriminal groups or other financially motivated groups as top candidate threat actors because compromising AI integrity offers minimal financial returns compared to attacks like stealing the model weights or traditional cybercrime like ransomware or data theft. Furthermore, the technical sophistication required to execute such attacks is not justified by the limited financial payoff.[41]

capabilities, enabling a large number of moderately skilled attackers to pose threats that previously required nation-state resources. This shift toward "quantity over quality" in offensive cyber capabilities may increase the aggregate risk from lower-resourced threat actors, even if individual campaigns remain less sophisticated than nation-state operations.

[41] That said, there are some potential financial motivations for AI integrity attacks, though we assess these as insufficiently lucrative given the technical complexity and risk involved. For instance, if hedge funds and financial institutions rely on LLMs for economic analysis and trading signals, an attacker could poison training data or fine-tuning processes to introduce systematic biases such as artificially bullish sentiment toward specific securities or sectors. The attacker could then take positions in those assets before the biased recommendations propagate through the market, profiting from the resulting price movements (a form of front-running or market manipulation). However, such schemes face significant barriers: the unpredictability of LLM influence on actual trading decisions, regulatory scrutiny of unusual market activity, the need to coordinate attack timing with position-taking, and the substantial technical effort required relative to simpler financial crimes.

| Threat Actor Profiles | | | | | |
|---|---|---|---|---|---|
| **Threat Actor** | **Motivation** | **Technical Capability** | **Resource Constraints** | **Access Vectors** | **Consequences if Caught** |
| **Nation State** | Geopolitical competition, intelligence collection, sabotage | Very high—zero-days, sophisticated tradecraft | Minimal budget constraints | Network intrusion, insider recruitment, supply chain compromise, social engineering | Diplomatic costs, potential sanctions |
| **Insider Threat** | Ideology, foreign recruitment, personal grievance | High—legitimate access, system knowledge | Limited to individual access and expertise | Direct internal access to training infrastructure and data | Criminal prosecution |
| **Misaligned AI** | Pursuing emergent goals from training | Varies by model capability | Constrained by deployment environment | Direct internal access to training infrastructure and data | Retraining, modification, or decommissioning |
| **Terrorist** | Advancing political agenda, gaining military advantage | Moderate—limited sophisticated capabilities | Significant budget and expertise constraints | Public dataset poisoning | Military retaliation, criminal prosecution |

***Table 2. Summary of threat actors, their capabilities, motivations, and constraints relevant to AI integrity attacks.***

# 3 | Four Approaches to Preserving AI Integrity

Successfully defending against the threats outlined above requires implementing proactive defenses that address vulnerabilities at every stage of the AI lifecycle, from development to deployment. **We identify four complementary strategies for preserving AI integrity:**[42]

1. **AI infrastructure security**: protects the systems, networks, and processes used to develop and deploy frontier AI systems, preventing integrity attacks *before they occur*. This approach focuses on 3 areas: preserving model weight integrity, preserving training data integrity, and securing data filtering algorithms.
2. **Data auditing**: addresses the trustworthiness of the data by ensuring the data's quality, integrity, and provenance. This approach includes developing data filtering algorithms to detect and remove poisoned content as well as maintaining auditable records of what data was used and where it came from.
3. **Model auditing and evaluation**: identifies whether an AI system has been compromised after training is complete by evaluating the model itself. This approach includes behavioral evaluations that probe for suspicious patterns (black-box methods) and analysis of model weights to locate hidden functionality (white-box methods).
4. **AI control**: acknowledges that we may never achieve complete certainty that an AI system remains uncompromised. AI control methods focus on detecting and blocking malicious behavior during deployment to enable the safe and productive use of a potentially untrusted model. This approach applies *zero trust* principles to AI systems: just as zero trust network architectures assume no implicit trust based on location or ownership and continuously authenticate users and devices, AI control assumes no implicit trust in model outputs and continuously monitors and validates AI behavior.[43] AI control methods include real-time monitoring, restricted autonomy, and isolation.

[42] Note that our four strategies for preserving AI integrity are not exhaustive. For example, one fruitful approach for reducing risk from compromised AIs is to procure AIs from multiple developers. This ensures that even if one AI developer is compromised the other AI developer's models are trustworthy. In practice, this could look like the government jointly procuring AI systems from OpenAI, Anthropic, and Google DeepMind. Then, government employees could use different AIs to check each other's work.
[43] U.S. Army Cyber Command, "Zero Trust."

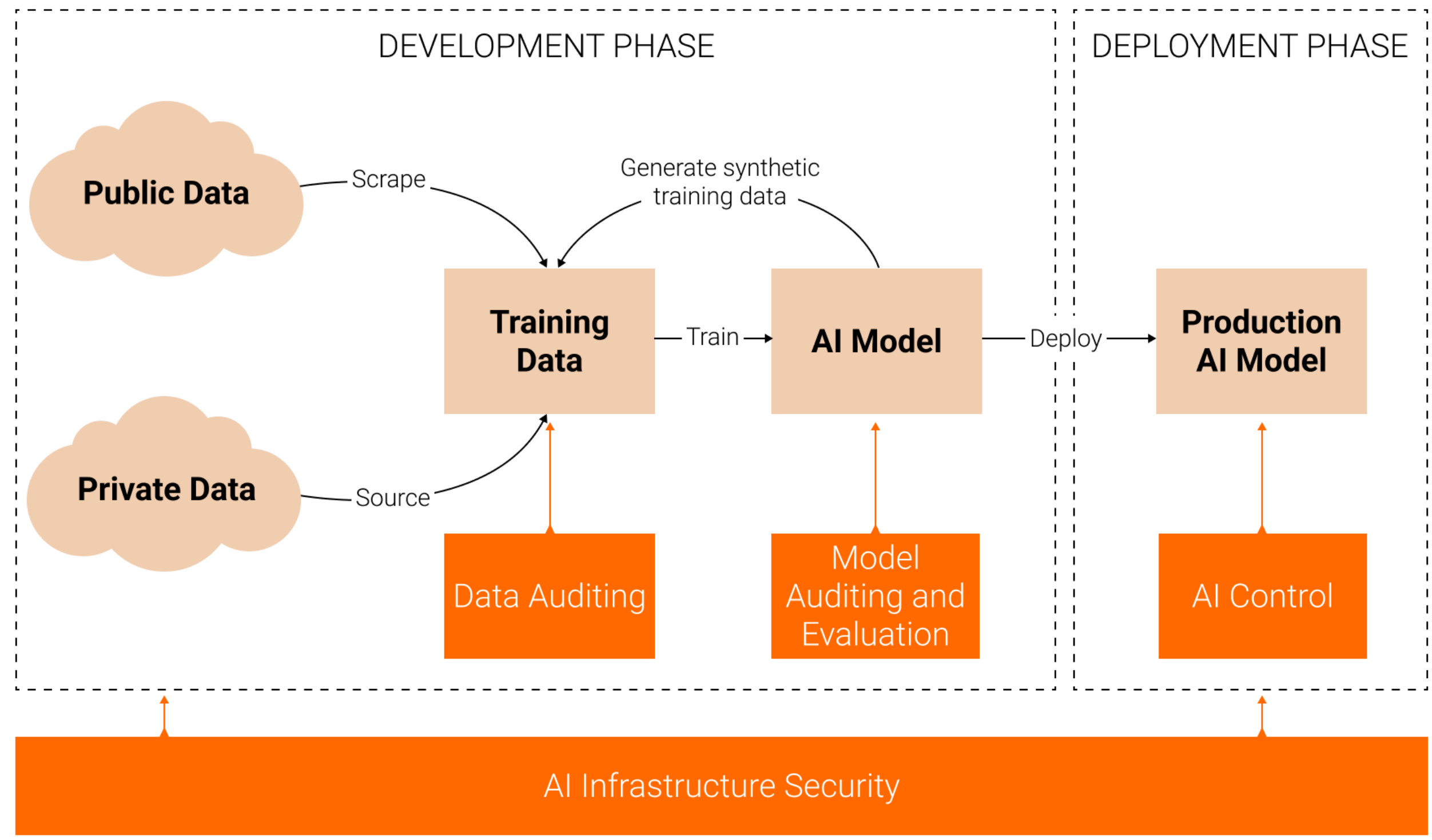

*Figure 4. Summary of the four primary approaches to preserving AI integrity.*

## Approach 1: AI Infrastructure Security

This first approach aims to prevent unauthorized access and modifications throughout the AI development and deployment lifecycle. AI infrastructure has a massive attack surface, so we focus on the components most important for preserving AI integrity: **model weight integrity**, **training data integrity**, and **security of data filtering algorithms**.

***Model weight integrity*** ensures that AI model weights remain free from unauthorized or secret modifications during both training and inference. In practice, this means preventing attackers from "swapping" legitimate model weights with poisoned versions.[44] Preserving model weight integrity requires addressing:

- **Inference integrity verification**: Periodically verifying that served models match reference model hashes. Ideally, this integrity verification could happen through a hardware-enforced channel. Verifying integrity during *training* is significantly harder, because frequent weight updates mean there is no stable reference hash to check against. One promising direction is developing methods for deterministic, replicable training runs that can be independently audited.[45] Further research is needed on integrity verification during training.

[44] Similarly, we may also be concerned about system prompt swap attacks. A malicious system prompt could include the password trigger for a backdoor, for example.

- **Model provenance**: Maintaining comprehensive audit trails requires storing snapshots at each training step.[46] Ensuring model provenance may involve storing a data structure containing some or all of the following information at each gradient step: pre/post-update weights, gradients, optimizer states, hyperparameters, loss values, random seeds, and training data batches. Unfortunately, this would likely create major storage bottlenecks. Further research is required to develop a scalable solution for maintaining model provenance.
- **Deployment controls**: Deploying new models should require strong organizational controls. Multi-party authorization should be required before production deployment, with independent verification that the deployed model matches authorized checkpoints. Logs should document who approved deployment, what testing occurred, and what the intended model behavior should be.

***Training data integrity*** ensures that training data remains free from unauthorized or secret modifications, as well as ensuring that the data is high-quality and free from poisoning. Preserving training data integrity requires addressing:

- **Tamper-resistant storage**: Implementing robust access controls limiting who can modify datasets, immutable storage systems[47] (e.g., write-once-read-many [WORM] drives,[48] blockchain-based storage[49]), multi-party authorization for dataset modifications, and periodic integrity checks via re-hashing.
- **Data redundancy**: Maintaining multiple copies of data across geographically separated locations to protect against accidental corruption and targeted attacks. These redundant data copies should ideally be tamper-proof and periodically integrity checked via hashing.
- **Transformation logging**: Recording all modifications, filters, and transformations applied to data. This creates an auditable record enabling post-incident investigation if compromised data is later discovered and helps identify which transformations may have introduced poison.

---

[45] Shavit, "What Does It Take to Catch a Chinchilla?"
[46] Training an AI model is a non-deterministic process, which makes model provenance auditing difficult. If auditors are unable to reproduce a training run, they may struggle to rule out foul play. Deterministic training warrants further research.
[47] Ahmad et al., "HardLog: Practical Tamper-Proof System Auditing Using a Novel Audit Device."
[48] Guardiola et al., "The Reverse File System: Towards Open Cost-Effective Secure WORM Storage Devices for Logging."
[49] Aniello et al., "A Prototype Evaluation of a Tamper-Resistant High Performance Blockchain-Based Transaction Log for a Distributed Database."; Putz et al., "A Secure and Auditable Logging Infrastructure Based on a Permissioned Blockchain."

- **Provenance metadata**: For pre-training data, recording stable source URLs and timestamps. For synthetic data, recording which model generated each sample (this is critical for preventing subliminal learning attacks[50] since such attacks work best when the model is trained on its own outputs[51]).
- **Third-party data verification**: Treating externally-sourced data (e.g., from Scale AI[52]) with heightened scrutiny. Further research is needed on methods for data providers to credibly assure procurers of data integrity, such as auditable chain-of-custody records or cryptographic provenance guarantees.

***Security of data filtering algorithms*** ensures that attackers are unable to compromise the confidentiality and integrity of data filtering algorithms. Preserving the *confidentiality* of data filtering algorithms makes it harder for attackers to iterate on the defender's defenses (i.e., if the attacker does not understand how the data filtering algorithms work, then they will have a harder time crafting poisoned data that bypasses those data filtering algorithms). Preserving the *integrity* of data filtering algorithms makes it difficult for attackers to sabotage the data filtering algorithms through "*algorithm swap attacks*," where the attacker replaces a legitimate data filtering algorithm with a more inaccurate one, or "algorithm degradation attacks," where the attacker directly tampers with the development of data filtering algorithms to make them inaccurate or unreliable.[53] Securing data filtering algorithms requires addressing:

- **Information compartmentalization**: Separating teams working on different filtering stages (e.g., one team detects spam, another handles steganography) to prevent insider threats from understanding the complete pipeline.[54]
- **Algorithm integrity controls**: Preventing algorithm swap or degradation attacks through multi-party authorization for updates and tamper-proof audit logs documenting all changes. Overall, the field of securing algorithmic secrets remains significantly underdeveloped and requires further research.

---

[50] Cloud et al., "Subliminal Learning: Language Models Transmit Behavioral Traits via Hidden Signals in Data."
[51] Subliminal learning occurs when a teacher model with some trait T generates innocuous data D unrelated to T. When a student model is trained on D, it also acquires trait T. However, subliminal learning only works if the student and teacher models are derived from the same base model. In the context of data poisoning defenses, this means frontier AI developers should treat data generated by the frontier model being trained with special scrutiny, as the subliminal learning effect may enable covert trait propagation.
[52] Scale AI, "Data Engine: Data Annotation, Collection, & Curation Platform."
[53] This could be considered a "meta poisoning attack." Rather than directly poisoning the frontier AI's training data, the attacker compromises the data used to train the filters themselves. By degrading the accuracy of these filters, the attacker creates a path to successfully poison the frontier model downstream.
[54] Information compartmentalization has downsides. It reduces productivity, which could result in lower-quality data filters. It also concentrates power among the few researchers with knowledge of the filtering pipeline.

- **Canary testing**: Intentionally planting known-malicious samples that legitimate data filters should catch. If these known-malicious samples bypass the filters, then the data filters may have been compromised.

---

Many of the security measures listed above impose significant productivity taxes, creating friction that prevents industry adoption. Research into making these implementations low-friction and developer-friendly is necessary for practical deployment. Additionally, due to the threat posed by insider threats, we recommend frontier AI developers strengthen insider threat programs.

## Approach 2: Data Auditing

The second approach addresses the quality trustworthiness of training data itself. Since AI systems learn their behaviors from training data rather than explicit programming, ensuring data integrity is critical to preventing compromised AI models. Data auditing measures include:

- **Data filtering algorithms**: Employing data filters to detect and eliminate poisoned data from training datasets.
- **Steganography detection**: Identifying malicious content hidden within seemingly benign training data, where backdoor triggers or biases are steganographically encoded to evade standard filtering techniques.[55]
- **Data attribution**: Frameworks for tracing specific model behaviors back to specific training data samples.[56]

Unfortunately, data auditing remains significantly underdeveloped. Training datasets often contain trillions of tokens sourced from diverse origins, meaning even *extraordinarily* accurate filters will still allow substantial amounts of poisoned data through—enough to compromise model behavior.[57] Moreover, attackers can craft poisoned data specifically designed to evade automated filtering (e.g., using subliminal learning techniques where malicious patterns are encoded in ways that appear innocuous to current detection algorithms[58]). As AI companies increasingly rely on training on synthetic data,[59] ensuring the integrity of these synthetic datasets should become a high priority.[60]

[55] Cloud et al., "Subliminal Learning: Language Models Transmit Behavioral Traits via Hidden Signals in Data."
[56] Berdoz et al., "Data Attribution in Large Language Models via Bidirectional Gradient Optimization."

We think some of the most important open problems in training data auditing are:

- **Detecting obfuscated poisoned data**. Techniques like Phantom Transfer produce poisoned datasets that evade even oracle-level defenses where the defender knows exactly what the attack is targeting.[61] To the best of our knowledge, there are no techniques for detecting inductive backdoors, phantom transfer attacks, or subliminal learning attacks.
- **Targeted poisoned data retrieval**. If defenders know the specific backdoored behavior they are looking for (e.g., insecure code generation or pro-adversary sentiment), can they reliably locate the responsible training samples? This is a more tractable variant of the general data filtering problem because the space of catastrophically bad behaviors may be small.
- **Scaling data filters without sacrificing precision**. Many data filtering techniques achieve strong detection rates in controlled settings but become impractical at scale because they filter out too much benign data (i.e., high false positive rate). It would be valuable to develop filters that maintain both high true positive rates and low false positive rates across trillion-token datasets.

---

[57] Souly et al., "A small number of samples can poison LLMs of any size."; To illustrate why data filters require extraordinary accuracy, let's assume a data filtering algorithm achieves a 99.9% true positive rate in filtering out poisoned data. Based on Anthropic's recent paper on data poisoning, 250 documents averaging 1150 tokens each (250*1150=287K total poisoned tokens) sufficed to backdoor a model. Consider a frontier model with 5T trainable parameters (similar to public estimates of GPT-4.5's size) trained under Chinchilla-optimal scaling (20 tokens per parameter). Such a model uses 5T*20=100T total training tokens. With a 99.9% true positive rate data filter, an attacker would need 287M poisoned tokens to successfully backdoor the model—merely 0.00029% of the training corpus.

Furthermore, the number of poisoned samples required to backdoor a model is sensitive to filter accuracy. At 99% true positive rate, attackers only need 29M poisoned tokens (0.000029% of corpus).

To complicate matters further, the above analysis assumes that the false positive rate is zero. However, in realistic scenarios, the data filters would be "catching" significant amounts of benign data (i.e., a high false positive rate). The upshot is that even pretty good data filters will hurt downstream model capabilities.

[58] Cloud et al., "Subliminal Learning: Language Models Transmit Behavioral Traits via Hidden Signals in Data."

[59] xAI, "Grok 4."

[60] More specifically, we must ensure the base model generating synthetic data is not itself compromised. This expands the attack surface considerably: compromised models could propagate secret loyalties to their successors through the synthetic training data they produce.

[61] Draganov et al., "Phantom Transfer: Data-level Defences are Insufficient Against Data Poisoning."

## Approach 3: Model Auditing and Evaluation

The third approach addresses auditing the AI model itself[62]—examining its parameters and behavior to identify tampering, backdoors, or unauthorized modifications. There are two primary methods, distinguished by whether evaluators examine the model's behavior or its internal workings:

1. **Black-box evaluation** probes how the model responds to carefully designed inputs (without access to model weights). Researchers test the model with various prompts and scenarios to identify suspicious patterns—for instance, whether the model exhibits systematic bias toward CCP narratives when asked about sensitive geopolitical topics. Unfortunately, black-box evaluations are inadequate and are unlikely to uncover basic backdoors or sophisticated secret loyalties.[63]
2. **White-box evaluation** examines the model's internal workings directly (with access to model weights). Researchers analyze the model's parameters and internal representations to identify anomalies that might indicate tampering. One key technique is mechanistic interpretability, a field focused on reverse-engineering AI models to provide insight into how and why an AI model exhibits particular behaviors. While more technically demanding, white-box methods can often detect backdoors that black-box testing misses.[64]

Model auditing should occur throughout the development process, not just after final training. Frontier AI developers should implement automated evaluation suites that run black-box and white-box tests at every model checkpoint—after pre-training, during post-training phases, and at intermediate stages. This requires robust model fingerprinting and version control: maintaining cryptographically signed records of model checkpoints and their associated metadata (training data sources, hyperparameters, evaluation results) to verify that each model version matches its intended state and to detect unauthorized modifications between checkpoints.

Both black-box and white-box evaluation techniques face significant challenges. Frontier AI models contain hundreds of billions or trillions of parameters, making comprehensive audits computationally expensive. Moreover, adversaries could design backdoors specifically to evade detection methods. This creates an ongoing adversarial dynamic between attackers and defenders, where each advance in detection capabilities may be countered by more sophisticated attack techniques.

---

[62] Approach 3 (model auditing and evaluation) benefits from access to the training data, so in most circumstances, it is advantageous to combine approaches 2 (data auditing) and 3 (model auditing and evaluation).
[63] Casper et al., "Black-Box Access is Insufficient for Rigorous AI Audits."; MacDiarmid et al., "Simple Probes Can Catch Sleeper Agents."
[64] *Ibid.*

**Ultimately, pure black-box evaluation methods seem insufficient for national security needs, as they are unlikely to uncover some integrity violations.[65] Defense efforts should prioritize multi-layered strategies, such as combining white-box evaluation techniques with training data audits.**

## Approach 4: AI Control

The fourth approach acknowledges that we may never achieve complete certainty that an AI system's integrity has been preserved. This raises an important question: *can we still safely and productively deploy potentially compromised models?* Fortunately, AI control offers an answer. **AI control is a research field focused on detecting and blocking malicious behavior during deployment to enable the safe deployment of a potentially compromised model**.[66] Rather than requiring absolute certainty about a model's integrity, these techniques enable organizations to benefit from AI systems while constraining their potential for harm. This approach is reflected in the White House AI Action Plan, which calls for investment in "AI control systems" as part of its strategy for AI security.[67]

AI control extends *zero trust* principles from network security to AI systems. Traditional network defenses assumed everything behind a firewall was safe; zero trust instead requires continuous verification of every user and device.[68] Similarly, AI control rejects the assumption that a model passing pre-deployment evaluation can be trusted thereafter, instead requiring ongoing validation of AI behavior throughout operation. AI control techniques include:

- **Real-time monitoring**: Logging all AI system interactions and outputs for human review, with particular scrutiny for sensitive operations
- **Limited autonomy**: Restricting AI systems to narrow, well-defined tasks rather than granting broad decision-making authority
- **Sandboxing**: Deploying AI systems in isolated environments where potential damage from compromised behavior can be contained
- **Chain-of-thought monitoring**: Inspecting a model's intermediate reasoning steps to detect deceptive or misaligned intent before final outputs are executed[69]

An added benefit of AI control is that if malicious behavior is detected during deployment, the compromised models can be decommissioned. While the previous three approaches to preserving integrity provide assurance during *development*, approach four (AI control) provides assurance during *deployment*. For a comprehensive discussion of open problems in AI control research, see Greenblatt.[70]

[65] *Ibid.*
[66] Greenblatt et al., "AI Control: Improving Safety Despite Intentional Subversion."
[67] The White House, "Winning the Race: America's AI Action Plan."
[68] U.S. Army Cyber Command, "Zero Trust."
[69] Delaney et al., "Policy Options for Preserving Chain of Thought Monitorability."

## A Defense-in-Depth Strategy

Currently, all four approaches remain significantly underdeveloped. AI infrastructure security involves challenging engineering problems, and current solutions potentially impose substantial operational costs that can slow AI development cycles. Data auditing techniques struggle with detecting poisoned content hidden through steganographic techniques. Model auditing methods struggle to identify backdoors embedded within trillions of parameters. AI control remains the least developed approach, with significant open questions about implementation and effectiveness.

**A defense-in-depth strategy requires investment across all four approaches simultaneously.** No single method provides complete protection; only by layering these complementary defenses can we build integrity-preserving AI systems.

[70] Greenblatt, "An Overview of Areas of Control Work."

# 4 | Policy Recommendations for AI Integrity

**Preserving AI integrity cannot be solved by industry alone.** Frontier AI companies face a threat landscape that demands capabilities beyond typical corporate security. More specifically, defending against nation-state intelligence services and insider threats requires specialized expertise that most companies lack. Moreover, the technical problems of detecting backdoors in trillion-parameter models and preventing tampering across complex development pipelines remain largely unsolved. While AI companies employ talented security teams, they are constrained by limited resources, competing priorities, and the need to maintain rapid development cycles in a competitive market.

**The US government is uniquely positioned to address these gaps through strategic investment and coordination.** Federal agencies possess decades of experience countering nation-state adversaries and insider threats. Research organizations like Intelligence Advanced Research Projects Activity (IARPA) and the National Institute of Standards and Technology (NIST) have proven track records of mobilizing cutting-edge technical solutions to national security challenges. And government procurement power can create strong incentives for industry adoption of security best practices without imposing heavy-handed regulation that could stifle innovation.

We recommend four complementary actions to strengthen AI integrity:

1. **Organize government-led red team exercises** where teams from the Intelligence Community (IC) test infrastructure security, and private machine learning (ML) security companies attempt to poison frontier AI models.
2. **Create voluntary NIST security frameworks** providing concrete technical guidance for protecting AI systems throughout their development lifecycle.
3. **Establish an AI Information Sharing and Analysis Center (AI-ISAC)** enabling mutual threat intelligence sharing between frontier AI developers and national security agencies.
4. **Launch new ARPA programs** hiring three program managers focused on model subversion detection, AI infrastructure security, and AI control systems.

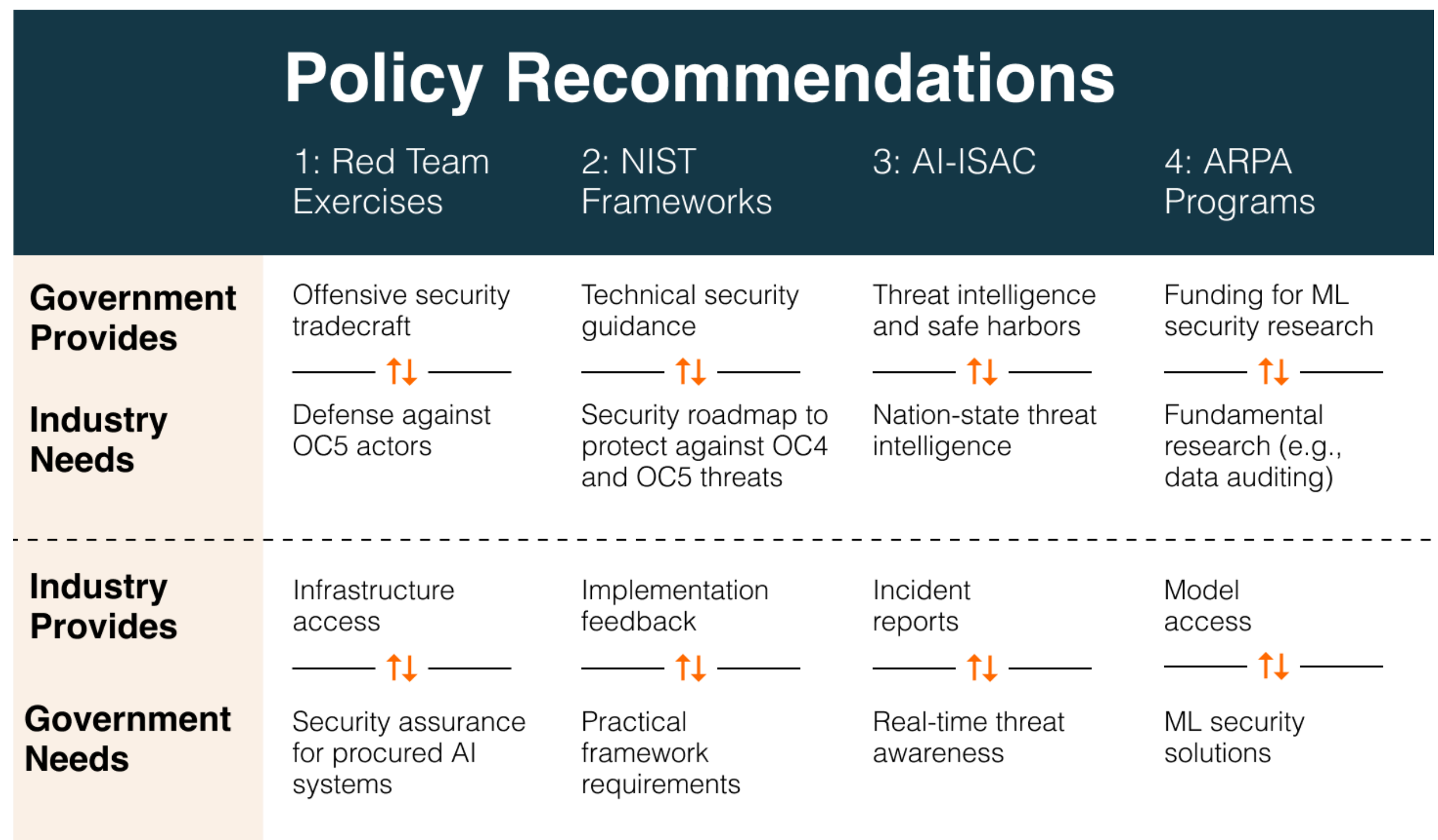

**Policy Recommendations**

| | 1: Red Team Exercises | 2: NIST Frameworks | 3: AI-ISAC | 4: ARPA Programs |
|---|---|---|---|---|
| **Government Provides** | Offensive security tradecraft | Technical security guidance | Threat intelligence and safe harbors | Funding for ML security research |
| **Industry Needs** | Defense against OC5 actors | Security roadmap to protect against OC4 and OC5 threats | Nation-state threat intelligence | Fundamental research (e.g., data auditing) |
| **Industry Provides** | Infrastructure access | Implementation feedback | Incident reports | Model access |
| **Government Needs** | Security assurance for procured AI systems | Practical framework requirements | Real-time threat awareness | ML security solutions |

***Table 3. Summary of AI integrity policy recommendations.***

## Recommendation 1: Organize Government-Led Red Team Exercises

**The US government should fund periodic red team exercises where expert teams attempt to compromise the integrity of AI systems at frontier companies.** This recommendation for government-led red teaming has clear historical precedent. In 1997, the DOD conducted Eligible Receiver 97 (ER97), a classified exercise in which an NSA Red Team used only publicly available tools and off-the-shelf equipment to simulate coordinated attacks against US critical infrastructure and military networks.[71] The exercise revealed that existing defenses were inadequate even against relatively unsophisticated attacks, exposing vulnerabilities across government and critical infrastructure. These findings directly catalyzed the formation of what would eventually become US Cyber Command.

The proposed AI integrity red team program seeks to replicate this model, using government-led adversarial exercises to reveal vulnerabilities in AI systems before real adversaries exploit them and to generate institutional momentum for defensive reforms.

[71] Martelle, "Eligible Receiver 97."

A successful AI integrity attack typically requires completing a kill chain with two distinct phases: (1) **compromising the infrastructure** through network intrusion, credential theft, or insider access, followed by (2) **exploiting the machine learning (ML) development process** through model sabotage or model subversion.[72] Traditional cybersecurity red teams excel at infrastructure compromise but typically lack expertise in ML attacks, while ML security specialists understand model-level vulnerabilities but may not possess nation-state-level operational tradecraft. Rather than attempting to source rare teams with deep expertise in both domains, this program decouples the exercises, allowing specialized teams to focus on their areas of strength while still testing the complete attack surface.

**To operationalize this program, the Department of Homeland Security (DHS), DOD, or IC should establish a red team program with two complementary components:**

1. **Infrastructure compromise exercises** led by IC teams focused on gaining unauthorized access to training data repositories, compute clusters, and development pipelines.
2. **ML-specific attack exercises** conducted by specialized security firms (e.g., Gray Swan AI, Palisade Research, Trail of Bits) that assume various levels of access[73] and attempt model subversion or sabotage attacks.[74]

The results of the exercises should be classified to prevent creating roadmaps for adversaries, though anonymized lessons should be shared across participating companies to raise the defensive baseline industry-wide.

**The program should begin with a pilot phase** targeting 1–2 major AI companies as voluntary participants to demonstrate value and refine methodology. This initial phase would establish

---

[72] Note that compromising the infrastructure is not necessary for integrity attacks that only involve poisoning public pre-training datasets.

Regarding pre-training data poisoning attacks, we are uncertain how to red team such attacks. Pre-training attacks need to "survive" throughout the entire development lifecycle—that is, poisoning a model in the pre-training phase might not work because the post-training phase may result in the poisoned behavior being "trained away." If the pre-training attack is successful and it is difficult to unlearn the poisoned behavior, then the frontier developer may have to restart their training run, which would be financially catastrophic.

[73] Davidson, "ML Research Directions for Preventing Catastrophic Data Poisoning."

[74] ML attack exercises should target dedicated test infrastructure (isolated compute clusters, replica training pipelines, representative checkpoints) rather than production systems. Attacking an actual training run risks irreversible model corruption and could cost millions in wasted compute (we acknowledge this approach trades some realism for safety, and we are not confident whether targeting test infrastructure is sufficiently realistic). These ML attack exercises should follow a progressive difficulty structure: initial rounds grant red teams maximal affordances (full system access, knowledge of defenses, extended timelines), then systematically reduce advantages in subsequent rounds. This reveals the minimum defensive posture required for blue teams to successfully detect and block attacks, informing practical security requirements for frontier AI development.

baseline procedures, identify operational challenges, and build trust with frontier AI developers before scaling.

Following a successful pilot completion, the program should expand through a phased approach. Year two could add 2–3 additional voluntary participants, incorporating lessons learned from the pilot phase. By year three, participation could become mandatory for all frontier AI developers holding federal contracts exceeding some dollar-value threshold or those designated as critical infrastructure. This timeline provides sufficient runway for companies to strengthen their security posture while ensuring comprehensive coverage across the frontier AI ecosystem within a reasonable timeframe.

## Recommendation 2: Create Voluntary AI Security Frameworks

**NIST, in collaboration with the Center for AI Standards and Innovation (CAISI), should develop voluntary security standards specifically designed for frontier AI developers.** These standards should provide concrete technical and organizational guidance focused on two security pillars: **confidentiality** (protecting model weights, training data, and algorithmic innovations) and **integrity** (ensuring AI systems remain free from unauthorized modifications or secret loyalties). By establishing clear standards rather than mandatory auditing requirements, this approach allows companies to adopt protections appropriate to their risk profile while creating benchmarks for government procurement.

Several recent initiatives point toward comprehensive AI security frameworks, such as The White House AI Action Plan's call for securing AI infrastructure and promoting secure-by-design AI technologies, and the SL5 Task Force's development of standards for protecting AI model weights against top nation-state threat actors.[75] **Building on these efforts, we propose that integrity protections be integrated into these standards rather than treated as a separate standard.** A comprehensive approach addressing confidentiality and integrity together provides clearer guidance for industry and avoids fragmenting security requirements across multiple competing frameworks.

Following the RAND security level (SL) framework,[76] these standards should define three progressive security levels calibrated to threat actor capability:

- **Level 1 (SL3 Integrity Protections)**: Defends against cybercrime syndicates and insider threats, including world-renowned criminal hacker groups, well-resourced terrorist organizations, and disgruntled employees. These actors operate with budgets up to $1 million and teams of roughly 10 experienced professionals.

[75] The White House, "Winning the Race: America's AI Action Plan."; SL5 Task Force, "SL5 Task Force - Security Level 5 for Frontier AI."; Grunewald and Gershovich, "Accelerating AI Data Center Security."
[76] Nevo et al., "Securing AI Model Weights: Preventing Theft and Misuse of Frontier Models."

- **Level 2 (SL4 Integrity Protections)**: Defends against standard operations by leading cyber-capable institutions, including many foreign intelligence agencies and state-sponsored groups. These operations deploy approximately 100 individuals with diverse expertise (e.g., cybersecurity, human intelligence, physical operations) over roughly one year with budgets up to $10 million.
- **Level 3 (SL5 Integrity Protections)**: Defends against top-priority operations by the most capable nation-states. These operations can deploy 1,000 individuals over multiple years with budgets approaching $1 billion.

The standard should be developed in three phases. First, Level 1 protections should be established to protect against operational capacity 3 (OC3) threats. Second, the standard should be extended to Level 2 protections against operational capacity 4 (OC4) threats. Finally, Level 3 protections against operational capacity 5 (OC5) threats should be developed once the necessary technologies mature. This phased approach allows iterative refinement while acknowledging that defenses against OC4 and OC5 threat actors require additional R&D before becoming production-ready.

**The standards should be scoped to include AI development *and* deployment security**. In development security, the standard should address data provenance tracking, model versioning and fingerprinting, access controls and multi-party authorization, and insider threat monitoring. In deployment security, the standard should address model and system prompt attestation,[77] model spec attestation,[78] and model weight integrity verification.[79]

By establishing these standards through NIST, the government provides clear security guidance without imposing rigid mandates. Companies also gain a roadmap for building trustworthy systems, and government agencies gain assurance that procured AI systems meet appropriate security thresholds. In the end, the broader AI ecosystem benefits from shared security baselines that raise the defensive floor across the industry.

## Recommendation 3: Establish an AI Information Sharing and Analysis Center

**The DHS should establish an AI Information Sharing and Analysis Center (AI-ISAC) that enables threat intelligence sharing between frontier AI developers, data center operators,**

[77] I.e., proving to the user that they are interacting with the claimed model version and system prompt.
[78] OpenAI, "OpenAI Model Spec."; Anthropic, "Claude's Constitution." I.e., proving to the user that a given model was trained via a specific model spec or constitution.
[79] I.e., verifying that deployed model weights match the cryptographically signed hash of the authorized reference weights through periodic integrity checks.

**and national security agencies.** This partnership would enable mutual intelligence flows: government agencies would share threat actor profiles and attribution analysis linking attack patterns to specific adversaries, zero-day vulnerability intelligence affecting AI development infrastructure, and real-time alerts about ongoing operations targeting the AI sector, while companies would share incident reports, technical details of newly discovered attack vectors, and early warning indicators of coordinated campaigns. By coordinating threat intelligence across frontier AI developers, each company gains visibility into attack patterns and vulnerabilities that others have observed, strengthening collective defenses.[80]

The AI Action Plan includes a related recommendation, calling for establishment of an "AI Information Sharing and Analysis Center (AI-ISAC), led by DHS, in collaboration with CAISI at DOC [Department of Commerce] and the Office of the National Cyber Director, to promote the sharing of AI-security threat information and intelligence across US critical infrastructure sectors."[81] However, the AI Action Plan's proposed AI-ISAC only focuses on critical infrastructure sectors that *use* AI systems rather than the companies that *develop* frontier AI. Protecting the integrity of frontier AI development requires extending intelligence sharing to include both developers and infrastructure operators.

**The AI-ISAC should address not only traditional cybersecurity threats but also ML-specific attacks.** Traditional ISACs focus on information sharing of the cybersecurity threat landscape, but frontier AI developers face a different threat landscape.[82] Sharing information about attacks like data poisoning is not necessarily protected under existing antitrust safe harbors. Thus, the DHS must explicitly establish a safe harbor for frontier AI developers to share ML-specific attacks, and not classify these attacks as "competitively sensitive business information." Industry initiatives like the Frontier Model Forum's information sharing agreement[83] demonstrate that AI companies recognize the need for coordination, but these efforts lack government participation and the legal protections necessary for sharing sensitive competitive information.

**The AI-ISAC should consist of the frontier AI developers (OpenAI, Anthropic, Google DeepMind, Meta, xAI), data center operators (AWS, Azure, GCP, Oracle Cloud), and national security agencies (NSA, FBI, CISA, and ODNI).** This partnership should draw lessons from successful programs like the Defense Industrial Base Cybersecurity program[84] and Financial Services ISAC.[85]

---

[80] National Council of ISACs, "National Council of ISACs."
[81] The White House, "Winning the Race: America's AI Action Plan."
[82] The White House, "FACT SHEET: Executive Order Promoting Private Sector Cybersecurity Information Sharing."
[83] Frontier Model Forum, "FMF Information-Sharing."
[84] U.S. Department of Defense, "Defense Industrial Base Cybersecurity Strategy 2024."
[85] FS-ISAC, "Financial Services Information Sharing and Analysis Center."

## Recommendation 4: Launch New ARPA Programs for AI Integrity Research

Preserving AI integrity requires solving two distinct classes of technical challenges: (1) cybersecurity threats and (2) machine learning threats. While defending against standard actors relies on known engineering best practices, **protecting against top-tier nation-state operations (OC5) involves genuinely unsolved research problems.** For example, securing distributed training clusters against such threats requires fundamental breakthroughs in trusted system architecture. **Similarly, the machine learning dimension presents genuinely unsolved research problems.** Detecting poisoned data in trillion-token training datasets, identifying steganographically encoded data, and auditing frontier-scale models for hidden objectives all require research breakthroughs. **These problems are inherently high-risk, high-reward research challenges suited for Advanced Research Projects Agencies (ARPAs).**

**We recommend that ARPAs (e.g., DARPA, IARPA, HSARPA, ARIA) source three new program managers with proven track records in ML and security, each focused on a distinct approach to AI integrity.**

**Program 1: Model Subversion Detection**. This program focuses on developing techniques to audit training data and AI models for poisoning and secret loyalties, building on the success of IARPA's TrojAI program.[86] This program should be structured around blue-team versus red-team audits to mirror real-world adversarial situations.[87] Research areas may include:

- Identifying specific circuits or mechanisms in model weights and activations that exist if and only if a backdoor has been planted[88]
- Investigating how innocuous-seeming data can be used to instill secret loyalties[89]
- Developing "last-mover" advantages for blue teams, such as techniques to "train out" or override unknown backdoors.[90]

**Program 2: AI Infrastructure Security.** This program would focus on developing security architectures to protect the physical and digital infrastructure used to train and deploy frontier AI models. For a detailed discussion of research areas in AI infrastructure security, see the "Approach 1: AI Infrastructure Security" section of this report.

---

[86] Intelligence Advanced Research Projects Activity, "TrojAI: Trojans in Artificial Intelligence."
[87] Davidson, "ML Research Directions for Preventing Catastrophic Data Poisoning."; Draganov et al., "Phantom Transfer: Data-level Defences are Insufficient Against Data Poisoning."
[88] MacDiarmid et al., "Simple Probes Can Catch Sleeper Agents."; Yu et al., "Backdoor Attribution: Elucidating and Controlling Backdoors in Language Models."
[89] Betley et al., "Weird Generalization and Inductive Backdoors: New Ways to Corrupt LLMs."; Cloud et al., "Subliminal Learning: Language Models Transmit Behavioral Traits via Hidden Signals in Data."
[90] Davidson, "ML Research Directions for Preventing Catastrophic Data Poisoning."

**Program 3: AI Control Systems**. This program would focus on building technologies that enable getting useful work out of untrusted AI models: real-time monitoring systems that detect malicious behavior during deployment; least-privilege frameworks that limit AI access to only essential information;[91] and multi-model verification protocols where independent systems must agree before high-stakes actions.[92] For a detailed discussion of open research problems in AI control, see Greenblatt.[93]

---

We also recommend establishing a Bay Area office to enable recruitment of top ML and security talent who cannot relocate to the East Coast. Offering part-time program manager positions could also significantly expand the talent pool, as many top researchers cannot fully leave their primary positions but could contribute 1–2 days per week to high-impact government research programs. To maximize the impact of these ARPA programs, we suggest the programs directly partner with frontier AI developers. These partnerships could involve:

- Secondment programs where company researchers spend 6–12 months working on ARPA projects while maintaining industry affiliations
- Joint research agreements that allow frontier companies to contribute proprietary datasets and model access for security research
- Regular technical working groups that bring together government, industry, and academic researchers to share findings (with appropriate classification controls)

[91] Debenedetti et al., "Defeating Prompt Injections by Design."
[92] Kalai et al., "Consensus Sampling for Safer Generative AI."; Ron et al, "Galápagos: Automated N-Version Programming with LLMs."; Wei et al. "CORTEX: Collaborative LLM Agents for High-Stakes Alert Triage."
[93] Greenblatt, "An Overview of Areas of Control Work."

# 5 | Conclusion

The rapid integration of AI systems across government operations increases exposure to AI integrity threats. As adversaries develop both the capability and motivation to compromise frontier AI systems, establishing robust defenses becomes increasingly important.

The four recommendations outlined in this report provide a strategy that leverages the government's unique strengths without imposing regulatory burdens on industry. **By organizing government-led red team exercises, developing voluntary NIST security frameworks, creating formal threat intelligence sharing partnerships, and launching new ARPA research programs, the government can significantly strengthen AI integrity across both public and private sectors.**

The path forward requires partnership between government and industry, combining the government's counterintelligence expertise with industry's technical innovation. By implementing these recommendations, the United States can maintain its leadership in AI while ensuring that the systems underpinning critical national functions remain trustworthy and aligned with American interests. Many open questions remain about how best to tackle the problem of AI integrity, and **we therefore encourage the research community to contribute to this nascent field**.

# Acknowledgements

We are grateful to the following people for providing valuable feedback and insights: Abbey Chaver, Stefan Torges, Evan Miyazono, Tom Davidson, Houlton McGuinn, Erich Grunewald, Matt Hodak, Nitzan Shulman, and Matthew Margo. Special thanks to Ashwin Acharya for encouraging us to begin research in this area. Acknowledgment does not imply endorsement of all claims or conclusions in this report.

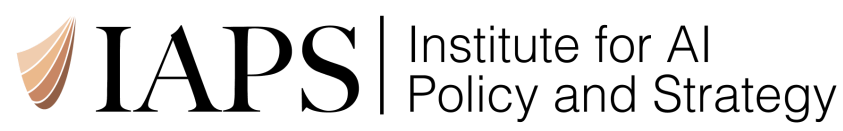